\documentclass{emulateapj}
\usepackage{apjfonts}

\usepackage{graphics,graphicx,rotating}
\usepackage{natbib}
\usepackage{xspace}
\usepackage{amssymb}
\usepackage{amsmath}
\usepackage{epstopdf}
\usepackage{subfigure}
\usepackage{color}
\usepackage{soul}

 \usepackage{textcomp}
\usepackage{gensymb}

\shorttitle{Shocks and Tides Quantified in \CIZSAUSAG}
\shortauthors{Molnar and Broadhurst}

\newcommand{\simless} 
     {\ensuremath{\lower 3pt\hbox{$\rlap{\raise5pt\hbox{$\char'074$}}\mathchar"7218$}}}
\newcommand{\simgreat}
     {\ensuremath{\lower 3pt\hbox{$\rlap{\raise5pt\hbox{$\char'076$}}\mathchar"7218$}}}

\newcommand{\simgt}{\lower.5ex\hbox{$\; \buildrel > \over \sim \;$}}
\newcommand{\simlt}{\lower.5ex\hbox{$\; \buildrel < \over \sim \;$}}

\newcommand{\nop}{{\noindent}}
\newcommand{\etal}{{\it et al.}}

\newcommand{\HE}{hydrostatic equilibrium}

\newcommand{\MSUNFOUR}{{$10^{\,14}\,$\ensuremath{\mbox{\rm M}_{\odot}}}\xspace}
\newcommand{\TMSUNFOUR}{{$\times 10^{\,14}\,$\ensuremath{\mbox{\rm M}_{\odot}}}\xspace}

\newcommand{\TMSUNFIVE}{{$\times 10^{\,15}\,$\ensuremath{\mbox{\rm M}_{\odot}}}\xspace}

\newcommand{\KMSEC}{{$\rm km\;s^{-1}$}\xspace}

\newcommand*{\ltsim}{\ {\raise-.75ex\hbox{$\buildrel<\over\sim$}}\ }
\newcommand*{\gtsim}{\ {\raise-.75ex\hbox{$\buildrel>\over\simsaus$}}\ }
\newcommand*{\proptosim}{\ {\raise-.75ex\hbox{$\buildrel\propto\over\sim$}}\ }

\newcommand{\MACH}{{\cal M}\xspace}
\newcommand{\MMACH}{{${\cal M}$}\xspace}

\newcommand*{\CHANDRA}{\emph{Chandra}\xspace}
\newcommand*{\XMM}{\emph{XMM-Newton}\xspace}
\newcommand*{\SUZAKU}{\emph{Suzaku}\xspace}

\newcommand*{\FLASH}{\emph{FLASH}\xspace}

\newcommand*{\COMPTONY}{Compton-$y$\xspace}

\newcommand*{\CIZSAUSAG}{CIZA J2242.8+5301\xspace}

\newcommand{\FIGURES}{./}

\begin{document}

\title{Shocks and Tides Quantified in the "Sausage" Cluster, \CIZSAUSAG, using 
N-body/\-hydro\-dynamical Simulations}

\author{
S. M. Molnar\altaffilmark{1} and T. Broadhurst\altaffilmark{2,3}
}

\altaffiltext{1}{Institute of Astronomy and Astrophysics, Academia Sinica, P. O. Box 23-141,
                      Taipei 10617, Taiwan}

\altaffiltext{2}{Department of Theoretical Physics, University of the Basque Country, Bilbao 48080, Spain}
                                            
\altaffiltext{3}{Ikerbasque, Basque Foundation for Science, Alameda Urquijo, 36-5 Plaza Bizkaia 48011, Bilbao, Spain}


\begin{abstract}
The colliding cluster, \CIZSAUSAG, displays a spectacular, almost 2 Mpc long 
shock front with a radio based Mach number $M\simeq 5$, that is puzzlingly large
compared with the X-ray estimate of $M\simeq 2.5$. 
The extent to which the X-ray temperature jump is diluted by cooler unshocked 
gas projected through the cluster currently lacks quantification. Thus, here we apply our 
self-consistent N-body/\-hydro\-dynamical  code (based on \FLASH) to model this 
binary cluster encounter. We can account for the location of the shock front and also 
the elongated X-ray emission by tidal stretching of the gas and dark mater between the 
two cluster centers. The required total mass is $8.9\,$\TMSUNFOUR with a 1.3:1 mass 
ratio favoring the southern cluster component. The relative velocity we derive 
is $\simeq 2500$ \KMSEC initially between the two main cluster components, 
with an impact parameter of 120 kpc.
This solution implies that the shock temperature jump derived from
the low angular resolution X-ray satellite \SUZAKU is underestimated by a factor of two, 
due to cool gas in projection, bringing the observed X-ray and radio estimates into agreement. 
We propose that the complex southern relics in \CIZSAUSAG, have been broken up as the southerly 
moving "back" shocked gas impacts the gas still falling in along the collision axis. 
Finally, we use our model to generate \COMPTONY maps to 
estimate the reduction in radio flux caused by the thermal Sunyaev-Zel'dovich (SZ) effect. 
At 30 Ghz, this amounts to 
$\Delta S_n = -0.072$ mJy/arcmin$^2$ and $\Delta S_s = -0.075$ mJy/arcmin$^2$ 
at the locations of the northern and southern shock fronts respectively. 
Our model estimate agrees with previous empirical estimates that have inferred the 
measured radio spectra can be significantly affected by the SZ effect, 
with implications for charged particle acceleration models of the radio relics.
\end{abstract}

\section{Introduction}
\label{S:Intro}

Massive pairs of colliding clusters display extreme physical effects, 
including huge X-ray shock fronts that are often traced by radio ``relics'' 
of large-scale diffuse synchrotron emission
(\citealt{Ensslin1999}; for a recent review see \citealt{FerettiET2012}).
Cluster Collisions are often recognized by a clear bimodal distribution of 
member galaxies and dark matter mapped by weak and strong lensing effects.  
The shocked gas in the iconic ``bullet cluster'' (1E0657-56) clearly implies that 
two massive clusters have just passed centrally through each other with a high 
relative velocity of $\simgreat \,3000\,$\KMSEC implied by the clearcut Mach cone 
found in its X-ray image \citep{MarkevitchET02}.
Velocities derived from X-ray observations using the shock jump conditions and 
N-body/hydrodynamical simulations support this interpretation 
\citep{MastBurk08MNRAS389p967,SpringelFarrar2007MNRAS380,MarkevitchET2004ApJ606,Molnar2013ApJ779},
 with a wide range of model velocities derived, 2700--4500 \KMSEC
 (for a recent review see \citealt{Molnar2015Frontiers}).
The Bullet cluster provides model independent support for the simple, collisionless assumption 
regarding dark matter, because the bimodal weak lensing pattern is coincident galaxy distribution 
 \citep{CloweET2006}, and hence, the dark matter must be essentially collisionless, like the member 
galaxies that have passed by each other during the first core passage.

The most massive cluster collisions have been uncovered in Sunyaev-Zel'dovich (SZ) 
sky surveys (ACT: Atacama Cosmology Telescope, \citealt{SifonET2013};
SPT: South Pole Telescope, \citealt{ReichardtET2012}
Plank Satellite, \citealt{Planck2011VIII}),
demonstrating that SZ selection favors cluster caught in collision when large 
columns of high pressure compressed gas are generated.
A prime example is the strong SZ source, ``El Gordo'', that is 
a binary colliding cluster discovered by ACT (ACT-CL J0102-4915, z = 0.87; \citealt{MenanteauET2012}).  
Using N-body/\-hydro\-dynamical simulations, we have shown its
cometary X-ray appearance can be readily reproduced by a collision that is not head on, 
generating two long, parallel tails of compressed gas, 
one of which has been pushed under gas pressure away 
from the most massive cluster component \citep{MolnarBroadhurst2015}
and the other tail is tidally compressed gas lying between the two cluster centers
(see also the simulations by \citealt{ZhangET2015}).
This compressed gas in ``El Gordo'' lies interior to several relatively small radio ``relics'' that 
appear to mark shock boundaries at the interfaces of gas belonging to each cluster lying at large radius.
It is difficult to detect merger socks because, although the pressure jump can be a factors of several,
but the gas density is very low.
However, a strong shock ($\MACH \,\simgreat \,3$) has been detected in ``El Gordo'' in the north-west
around a radio relic \citep{BotteonET2016,Basu2016ApJ829}.

Much larger radio relics have been uncovered in other colliding clusters
(e.g., 1RXS J0603.3+4214, the ``Toothbrush'' cluster, \citealt{OgreanET2013MNRAS433};
\CIZSAUSAG, the ``Sausage'' cluster, 
\citealt{AkamatsuET2015,OgreanET2014,OgreanET2013MNRAS429,WeerenET2010Sci}).
In \CIZSAUSAG a Mpc scale ``sausage'' shaped radio structure was found, which 
clearly corresponds to the gas collision shock front in the North, with its outer edge 
coincident with an X-ray temperature jump 
detected by \cite{AkamatsuET2015}.
This shock front is the most clearly defined example known, spanning an angle of $80\deg$.
Its highly polarized radio emission implies the presence of a cluster scale 
magnetic field of $\simeq 5\,\micro\,$G strength \citep{WeerenET2010Sci},
The sausage also shows another large, but more diffuse radio relic in the south that lies close,
but somewhat interior to the location of the southern X-ray shock claimed by \cite{AkamatsuET2015}.
This X-ray shock front is expected to be generated in the ``reverse'' direction, 
opposite to the faster moving infalling cluster that has moved through to the north 
after the first core passage 
\citep{RickerSarazin2001,Molnet2012ApJ748,MolnarBroadhurst2015}.
 No self-consistent hydrodynamical model has yet ben constructed for \CIZSAUSAG to date,
and here we will also discuss this reverse shock front based on our simulations. 

Several collision shocks have Mach number estimates from both radio and X-ray observations,
e.g., A2255; 1RXS J0603.3+4214, the ``Toothbrush'' cluster;
 \CIZSAUSAG, the ``Sausage'' cluster
\citep{AkamatsuET20161203058,OgreanET2013MNRAS433,AkamatsuET2015,
OgreanET2014,OgreanET2013MNRAS429,WeerenET2010Sci},
with a pattern emerging that the Radio estimates are generally significantly larger than the
X-ray estimates.  
In principle, the X-ray estimates simply relate the physical temperature change across the 
shock to the standard Rankine-Hugoniot jump conditions.
However,  X-ray observations detect projected temperatures, which are
derived from the observed spectrum as an integral through the entire cluster,
so that temperature estimates at any projected radius can comprise gas with a 
spread of temperatures. 
Also, in practice, the X-ray emission is weak in the outer regions where the 
shocks are often seen (e.g., \citealt{AkamatsuET2015}) 
requiring large area binning, and hence a large projected column of unshocked, 
cooler gas in each Xray bin may flatten the temperature jump, 
biassing the Mach number low. 
This problem is exacerbated by the presence of very hot and luminous shocked gas, 
so that a model of the gas distribution in the LOS is beneficial to help estimate the 
effects of projection and derive the physical quantities of interest, including the 
density and temperature distribution of the gas. 
Such a model can be derived using full N-body/\-hydro\-dynamical simulations 
constrained by observations (e.g., \citealt{MachadoET2015,MolnarBroadhurst2015}).

%
%
\begin{figure*}[t]
\includegraphics[width=1.0\textwidth]{\FIGURES/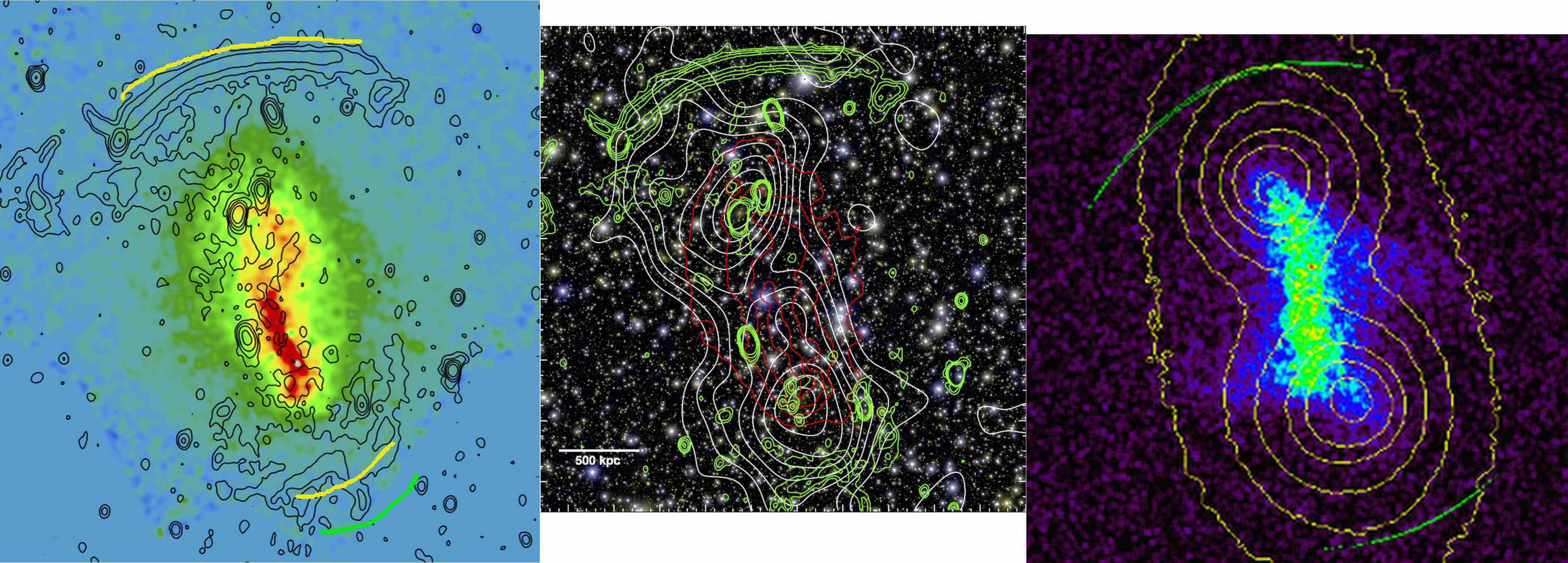}
\caption{
Images of \CIZSAUSAG based on multi-wavelength observations and our best model
derived using \FLASH simulations.
The first and second panels (from the left) show an image from \CHANDRA observations 
(Figure 1 from \citealt{OgreanET2014}) overlaid the shock positions
proposed by \cite{AkamatsuET2015} (yellow lines; 
the light green line is our estimate for the position of the southern shock)
and a Subaru image (gi color) with the galaxy number density contours (white contours), 
XMM-Newton X-ray luminosity (red contours), 
and Westerbork Synthesis Radio Telescope (WSRT) radio emission (green contours) 
overlaid (Figure 1 from \citealt{DawsonET2015}).
The third panel shows a simulated X-ray observation of our best model for 
\CIZSAUSAG (run P120V25) with projected mass distribution overlaid (yellow contours). 
The northern and southern shocks are marked with green contours.
The images are resized to the same physical scale 
(the length of the horizontal white bar in the middle panel is 500 kpc).
\vspace{0.3 cm}
\label{F:XRAYDMSHOCK}
}
\end{figure*} 

Here we generate a comprehensive set of self-consistent N-body/\-hydro\-dynamical 
simulations of binary merging clusters of galaxies 
constrained by observations of member galaxies, gas, and dark matter to 
model \CIZSAUSAG, and thus obtain the 3-dimensional (3D) structure of the 
shocks (e.g., gas density, temperature distribution). 
We prefer to base the gas dynamics on the \FLASH Adaptive Mesh refinement codes 
because of the presence of shocks that is not well captured by coarse grids or the
``smooth particle hydrodynamics'' approximation.

The structure of this paper is the following. 
In Section~\ref{S:SAUSAGE} we summarize results
from previous analyses of \CIZSAUSAG based on multifrequency
observations and numerical simulations.
We describe our simulation setup for modeling of \CIZSAUSAG
as a binary merger in Section~\ref{S:SIMULATIONS}.
Section~\ref{S:RESULTS} presents our results, 
and provide a physical interpretation of the morphology of multifrequency 
observations of \CIZSAUSAG. In this section we also discuss 
quantitatively the merging shocks in \CIZSAUSAG, 
as well as the biases in Mach numbers, derived from X-ray observations,
and in flux measurements of radio relics due to SZ contamination from shocks.
Section~\ref{S:CONCLUSIONS} contains our conclusions.

\section{\CIZSAUSAG: a merging cluster with the prototype of relics}
\label{S:SAUSAGE}

The most prominent radio relic is the northern relic in \CIZSAUSAG
(the ``Sausage'' Cluster: \citealt{WeerenET2010Sci}).
This is a bright radio relic, about 1.7 Mpc long, 
has a smooth bow or sausage shape
(thus called the ``sausage''), and it is only about
55 kpc wide, apparently delineating a merger shock.
Assuming a diffusive shock acceleration (DSA) model
(the most popular model to explain synchrotron radio emission from relics),
\cite{WeerenET2010Sci} derived a Mach number of 4.6.
On the opposite side of this merging cluster, on the southern part,
there are some scattered relics with irregular shape.

\CIZSAUSAG, at a redshift of $z = 0.19$ \citep{JeeET2015},
has been studied extensively.
Detailed spectroscopic observations of \CIZSAUSAG
were carried out by \cite{DawsonET2015}.
They found that the two subclusters have very similar redshifts,
suggesting that the collision is close to the plane of the sky.
\XMM and \CHANDRA observations of \CIZSAUSAG
were carried out to study the X-ray morphology and 
search for shocks and discontinuities in the X-ray emission by 
\cite{OgreanET2013MNRAS429} and \cite{OgreanET2014}.

Numerical simulations aiming to model the radio emission,
and constrain the initial parameters of the merging clusters were 
carried out by \cite{WeerenET2011MNRAS418}. 
They used N-body/\-hydro\-dynamical simulations
(\FLASH), but assumed fixed potentials for both subclusters. 
\cite{WeerenET2011MNRAS418} concluded that \CIZSAUSAG
is a binary merging cluster after the first core passage;
the collision is close to the plane of the sky ($\simless 10$\degree\
from the plane of the sky)
with a mass ratio of $\sim$2:1 (the northern subcluster being more
massive), and an impact parameter of $\simless \,400$ kpc.
Weak lensing observations were used to derive
the total mass of the system 
and the masses of the components of \CIZSAUSAG.
\cite{JeeET2015} found a total mass of M$_{200c,total} \sim 2.1\,$\TMSUNFIVE, 
a mass ratio $\sim 1:1$, and masses: 
M$_{200c,1} = 11.0_{-3.2}^{+3.7}\,$ \TMSUNFOUR and 
M$_{200c,2} = 9.8_{-2.5}^{+3.8}\,$\TMSUNFOUR 
(200c refers to a density 200 times the critical mass density).
\cite{OkabeET2015} derived a total virial mass of M$_{vir,total} \sim 1.9\,$\TMSUNFIVE
(M$_{200c,total} \sim 1.6\,$\TMSUNFIVE), a mass ratio $\sim 2:1$, and 
masses: M$_{vir,1} = 12.44_{-6.58}^{+9.86}\,$\TMSUNFOUR and 
M$_{vir,2} = 6.74_{-4.12}^{+7.64}\,$\TMSUNFOUR,
the southern subcluster found to be more massive.
Northern and southern shocks were found around the relics
with Mach numbers of $2.7_{-0.4}^{+0.7}$ and $1.7_{-0.3}^{+0.4}$ respectively, 
using \SUZAKU observations by \cite{AkamatsuET2015}.

Recently, the steepening of the spectra of the northern relic in 
\CIZSAUSAG, and the relic in 1RXS J0603.3+4214, the ``Toothbrush'' cluster,
was found \citep{StroeET2016}, which excludes simple DSA models.
Several solutions have been proposed to solve this problem
using the northern relic in \CIZSAUSAG as a prototype.
\cite{KangRyu2016,KangRyu2015ApJ809} proposed that fossil electrons 
(accelerated by the DSA mechanism) are 
reaccelerated as the shock runs them over;
\cite{FujitaET2016} suggested that cosmic rays are reaccelerated 
by turbulence generated behind shocks;
\cite{DonnertET2016} offered another explanation:
magnetic turbulence generated by the amplification of the magnetic field
is responsible for the particle reacceleration.
An alternative solution was suggested by \cite{BasuET2016AA591}:
the curved radio spectrum is a consequence of the SZ effect, 
which lowers the measured radio flux at the observed radio wavelengths
($ < 217$ GHz).

\section{Modeling \CIZSAUSAG using hydrodynamical simulations}
\label{S:SIMULATIONS}

The main goal of our project was to obtain a reasonable 
physical model for \CIZSAUSAG and study the merging shocks,
not to carry out a systematic search for the initial parameters
and determine their errors (that would require many more simulations).

\subsection{Details of the simulations, initial setup}
\label{SS:ICOND}

We model  \CIZSAUSAG using 3D self-consistent N-body/hydrodynamic 
numerical simulations of binary galaxy cluster mergers including dark matter and
intracluster gas.
The simulations have been carried out using the publicly available parallel Eulerian
adaptive mesh refinement (AMR) code \FLASH, 
developed at the Center for Astrophysical Thermonuclear Flashes
at the University of Chicago (\citealt{Fryxell2000ApJS131p273} 
and \citealt{Ricker2008ApJS176}).

We adopted a large box size, 13.3 Mpc on a side, for our simulations 
to capture the outgoing merger shocks.
We used our well established method to initialize and run the merging
cluster simulations 
\citep{MolnarBroadhurst2015,Molnar2013ApJ779,Molnar2013ApJ774,Molnet2012ApJ748}.
We reach the highest resolution, 12.7 kpc at the centers of the clusters
and at the merger shocks, as well as in the turbulent regions behind the shocks.
Our simulations were semi--adiabatic: we included only shock heating,
which is the most important non-adiabatic process for merging clusters.

%
%
\begin{figure}[t]
\includegraphics[width=.237\textwidth]{\FIGURES/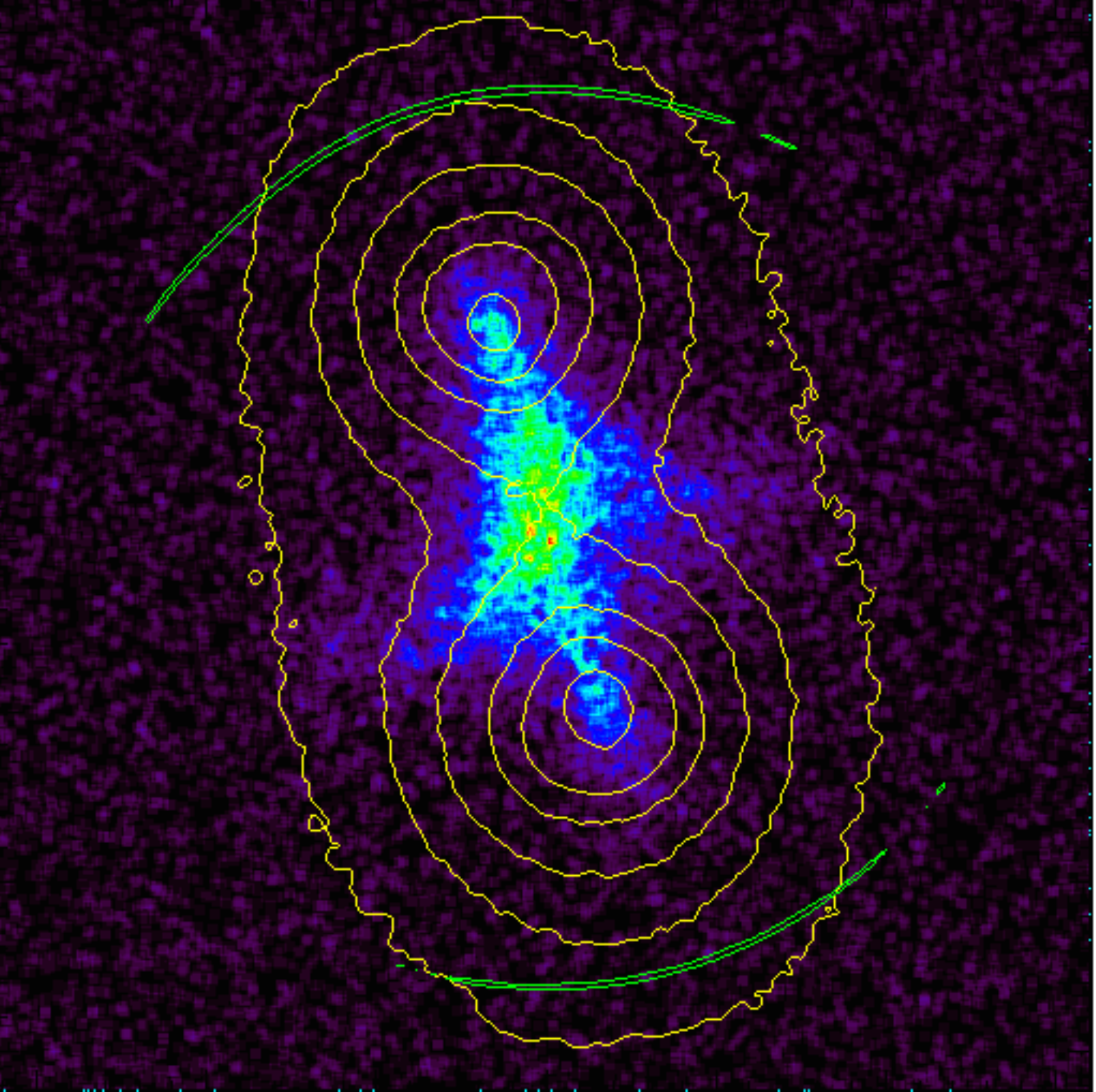}
\includegraphics[width=.237\textwidth]{\FIGURES/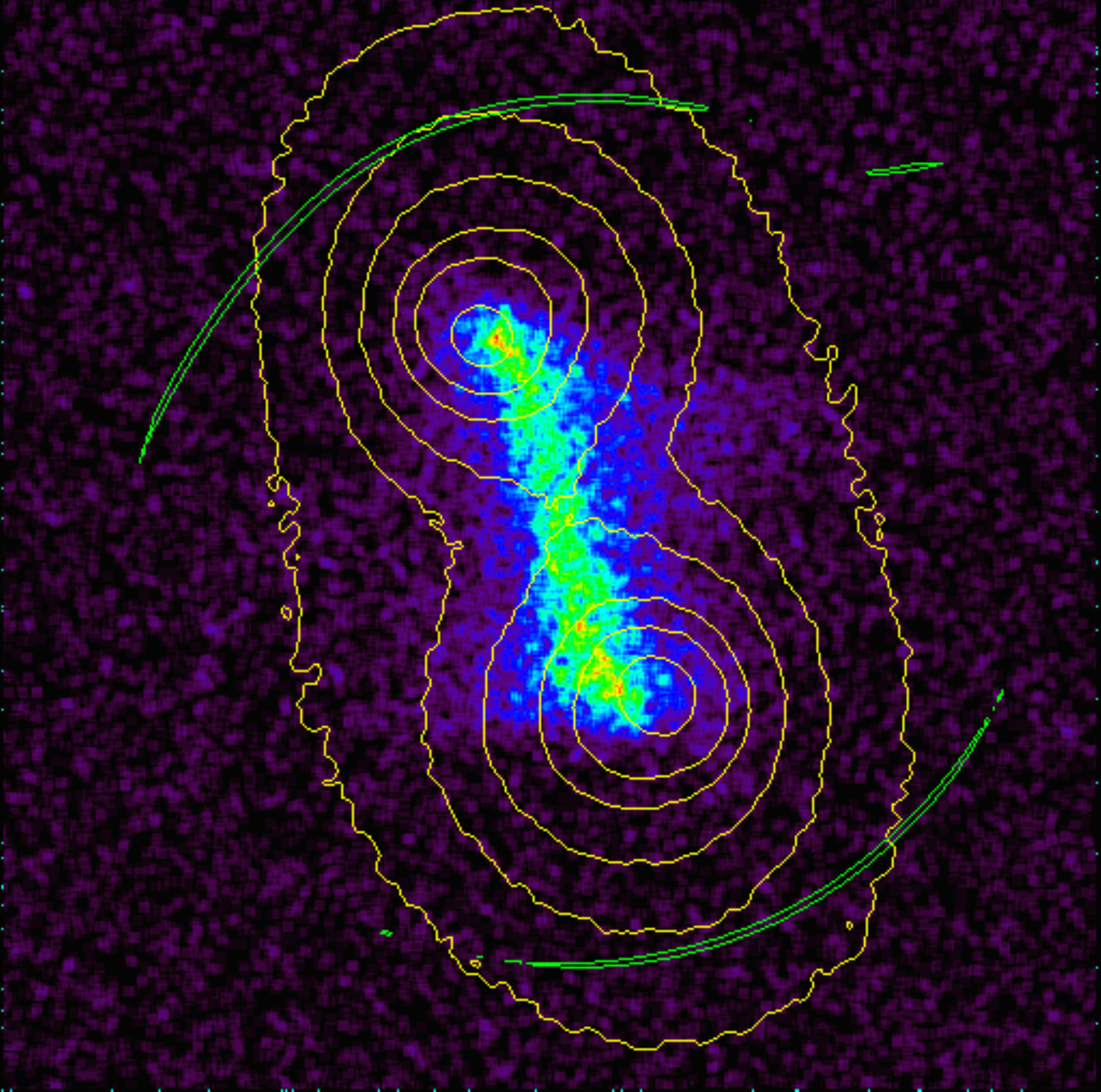}
\includegraphics[width=.237\textwidth]{\FIGURES/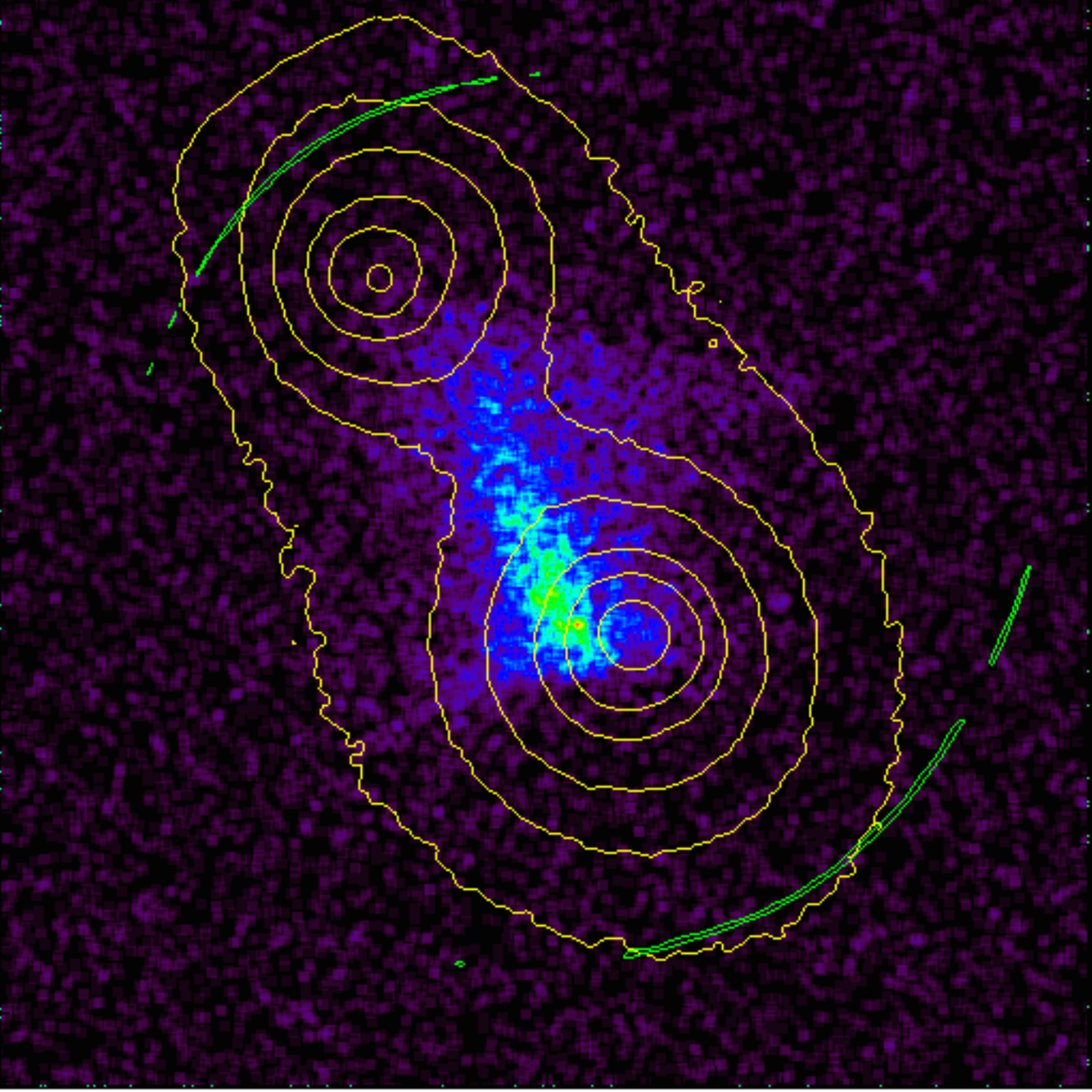}
\includegraphics[width=.237\textwidth]{\FIGURES/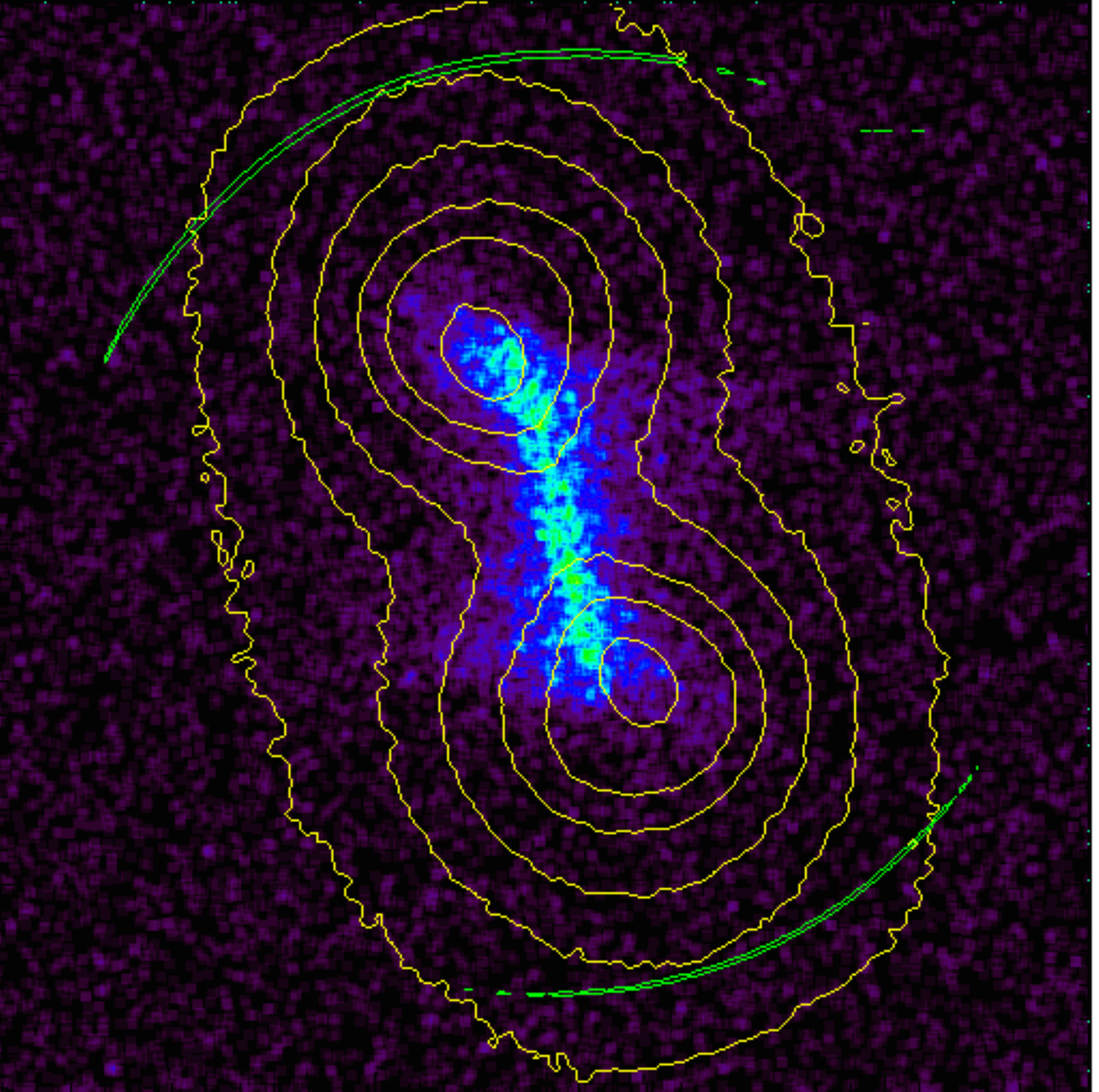}
\includegraphics[width=.237\textwidth]{\FIGURES/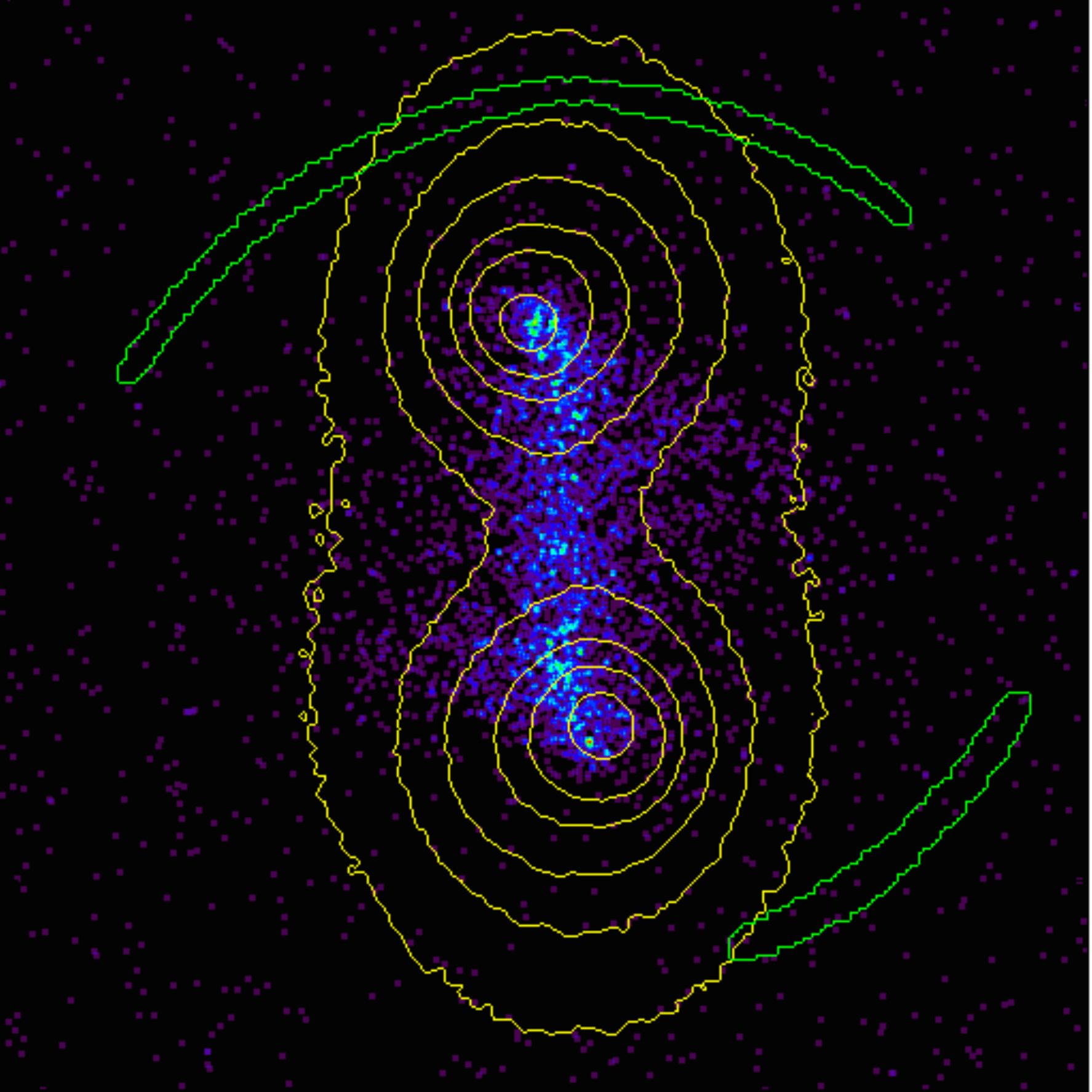}
\includegraphics[width=.237\textwidth]{\FIGURES/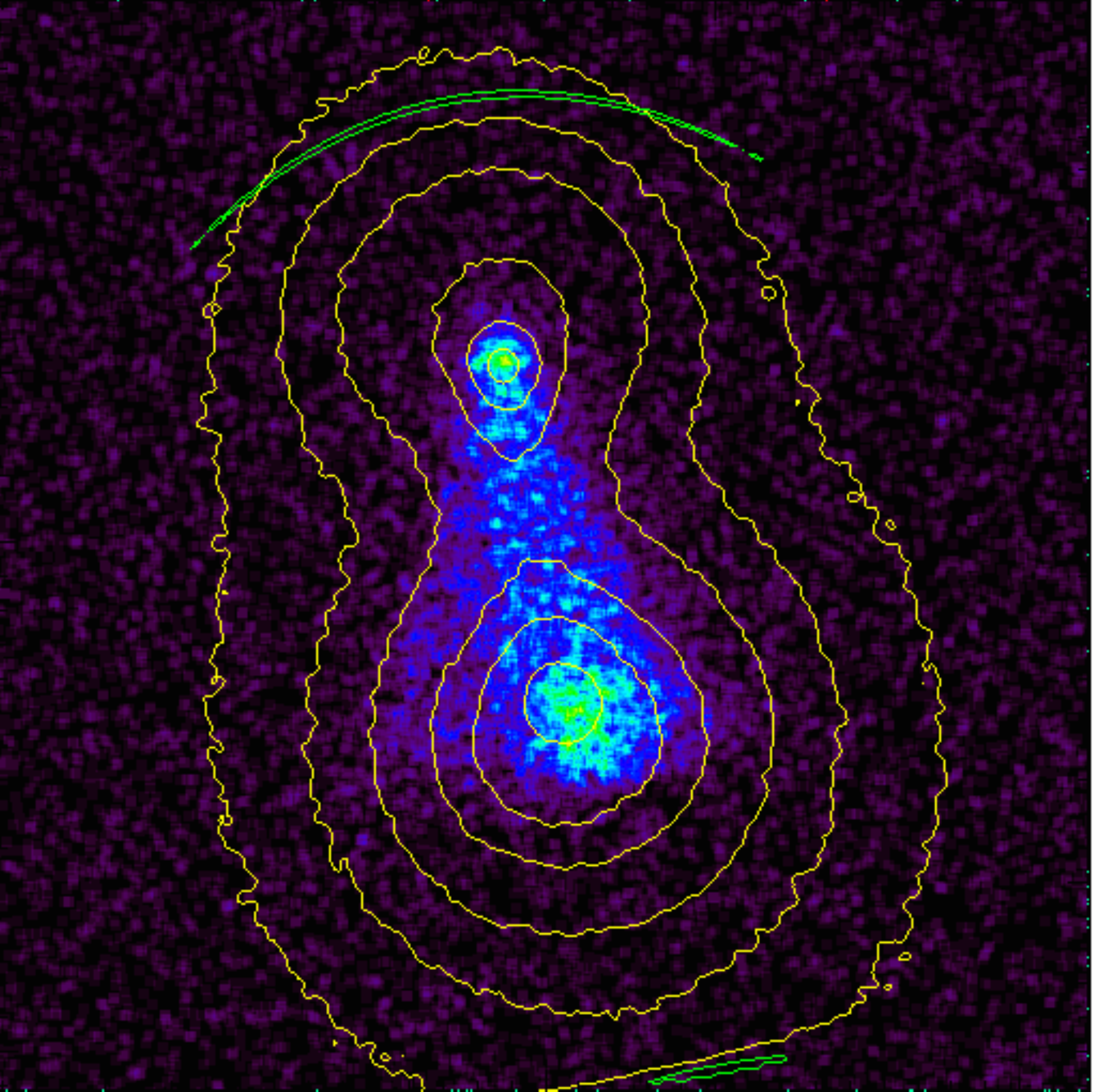}
\caption{
Models which do not do not match in detail \CIZSAUSAG.
The color coding is the same as in the third panel in Figure~\ref{F:XRAYDMSHOCK}.
The two panels in the first row show simulations with the 
same initial parameters as our best model, but different impact parameters: 
P = 50, and 180 kpc (runs P050V25 and P180V25),
the panels in the second and third row display images of runs with
varying mass ratios, impact parameters and infall velocities 
(runs P100V25, P150V20, P150V30, and  P200V18; see Table~\ref{T:TABLE1} 
for input parameters for our models).
\vspace{0.3 cm}
\label{F:XRAYBAD}
}
\end{figure} 

We assumed spherical clusters with cutoffs at the the virial radius ($r \le R_{\rm vir}$)
for the the dark matter and the intracluster gas.
Our initial conditions for the distribution of the dark matter was 
the NFW model \citep{NFW1997ApJ490p493}, 
\begin{equation} \label{E:NFW}
      \rho_{DM} (r) =  { \rho_s  \over x (1 + x)^2}
,
\end{equation}
\nop
where $x = r/r_s$, and $\rho_s$, $r_s = R_{vir}/c_{vir}$ are scaling parameters for the radius and the density,
and $c_{vir}$ is the concentration parameter, 
and for the gas density distribution we adopted a non-isothermal $\beta$ model,
\begin{equation}  \label{E:BETAMODEL}
      \rho_{gas}(r) =  { \rho_0  \over (1 + y^2)^{3 \beta /2} }
,
\end{equation}
\nop 
where $y = r/r_{core}$, is the scaling parameters for the radius,
$\rho_0$, is the density at the center, and $r_{core}$ is the scaling parameter
for the radius for the gas distribution.

We assumed \HE\ and derived the temperature from the equation of \HE\ adopting
 the ideal gas equation of state with $\gamma = 5/3$. 
We assumed $\beta=1$, suggested by cosmological numerical simulations
for the large scale distribution of the intracluster gas removing the 
filaments \citep{Molnet10ApJ723p1272}. 
We  used a gas fraction of 0.14, and represented the stellar matter in galaxies 
with collisionless particles, since galaxies can be considered 
collisionless for our purposes. 
The velocities of the dark matter particles were drawn from a Maxwellian distribution 
with a velocity dispersion as a function of distance from the cluster center obtained from 
the Jeans equation  assuming isotropic velocity dispersion \citep{LokasMamon2001MNRAS321}.
The distribution of the direction of the velocity vectors was assumed to be isotropic.
More details of the set up for our simulations can be found in \cite{Molnet2012ApJ748}.

%
%
\begin{figure*}[t]
\includegraphics[width=.325\textwidth]{\FIGURES/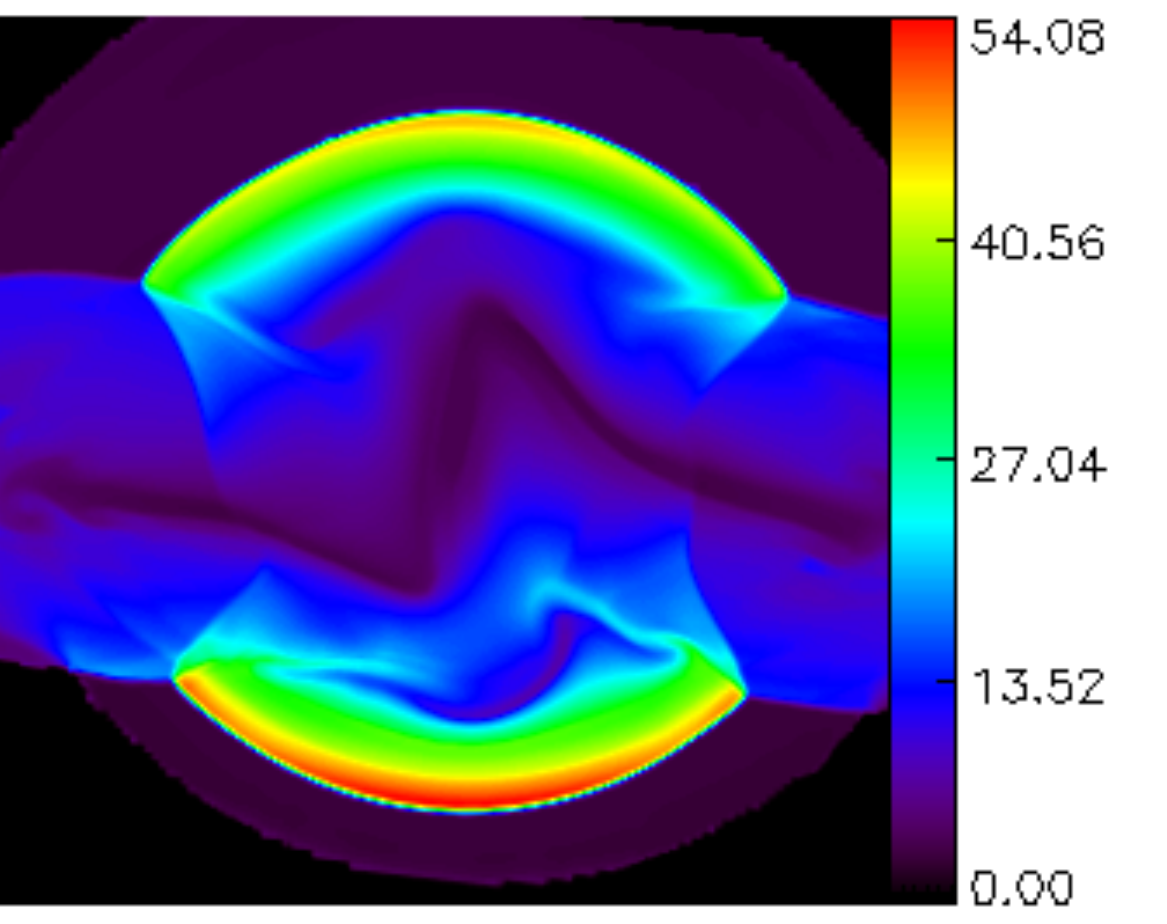}
\includegraphics[width=.325\textwidth]{\FIGURES/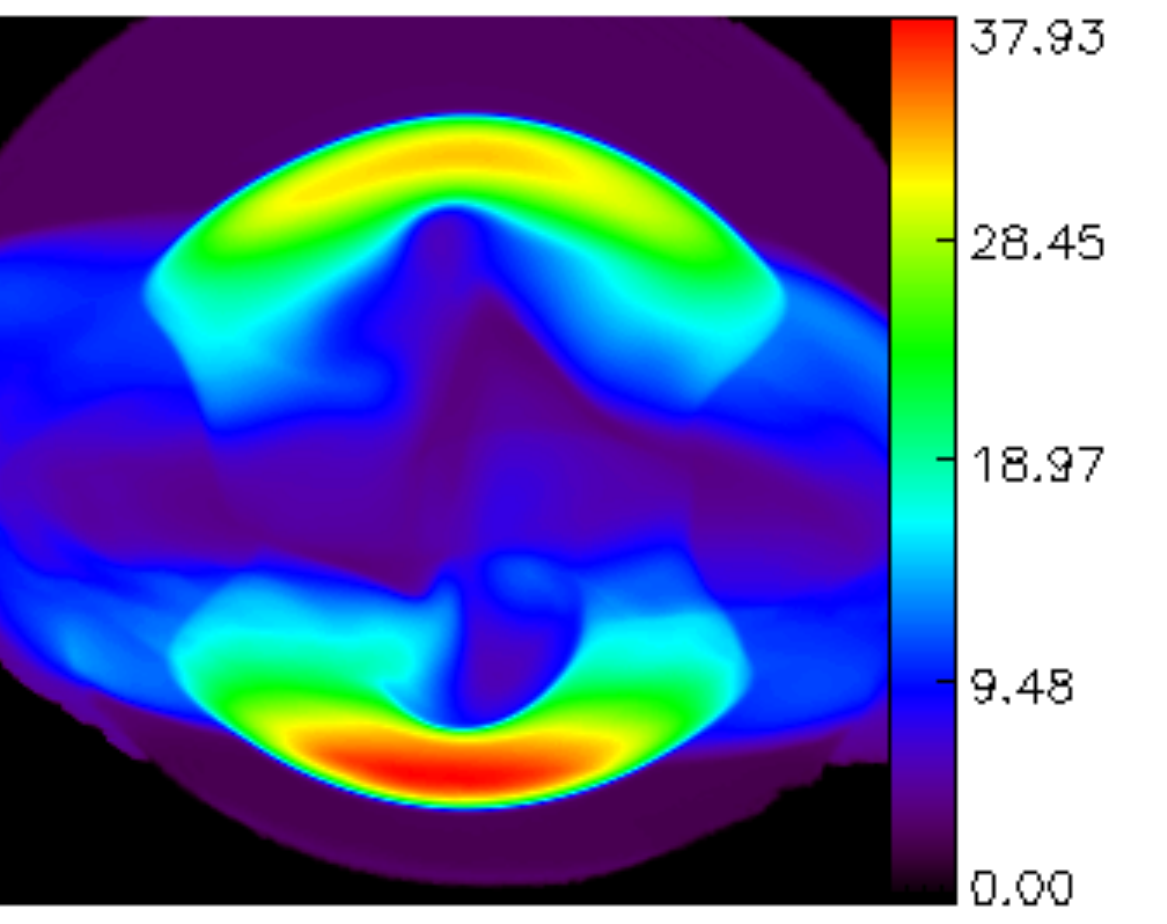}
\includegraphics[width=.325\textwidth]{\FIGURES/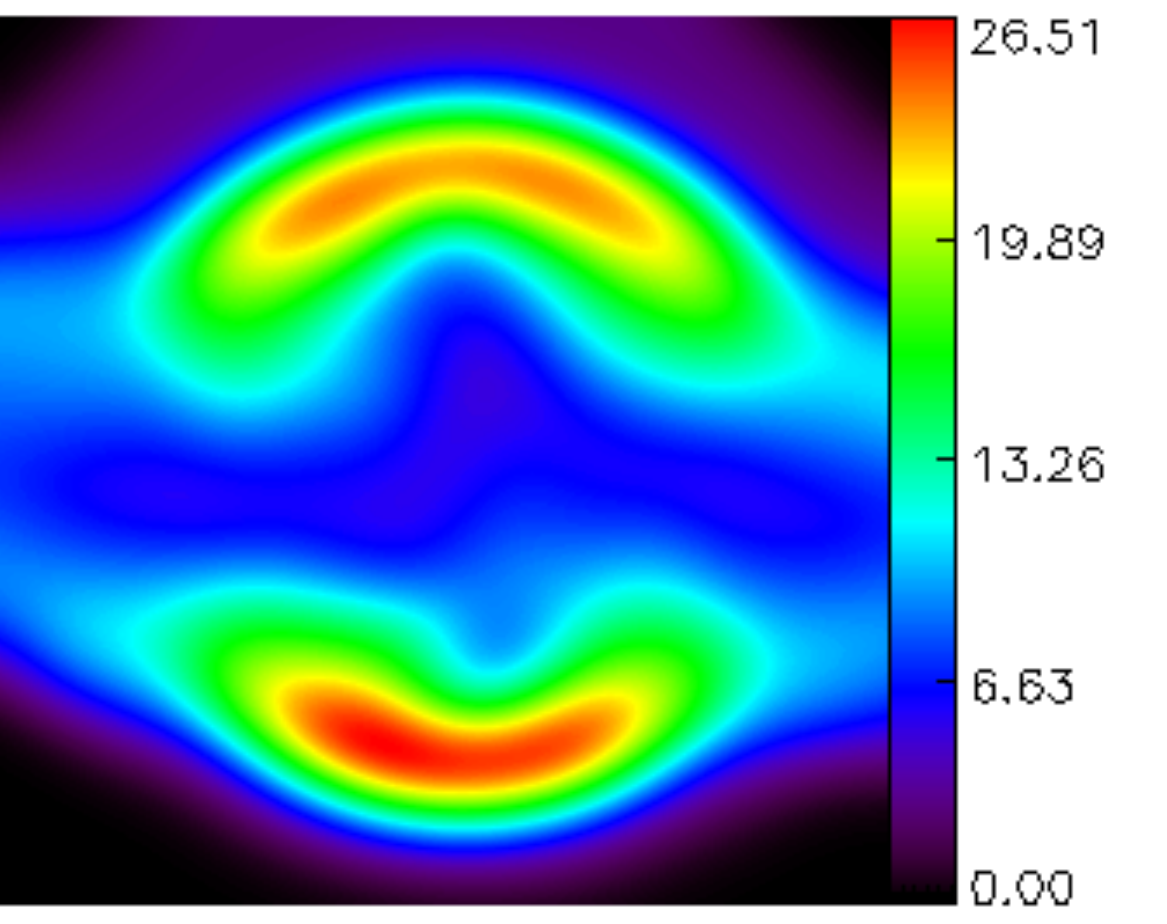}
\caption{
Temperature maps in units of keV from our best model for \CIZSAUSAG (run P120V25).
The high temperature regions (red/yellow) on the north and south mark the merging 
shocks moving outwards.
Note that the collision axis is aligned vertically to match the shock geometry.
{\it Left panel: }
Physical temperature, 2D cut through the main plane of the collision.
{\it Middle panel: }
Spectroscopic temperature \citep{MazzottaET2004}.
{\it Right panel: }
Mock \SUZAKU observation.
\vspace{0.3 cm}
\label{F:TEMPERSUZ}
}
\end{figure*} 

\subsection{\FLASH\ Runs}
\label{SS:RUNS}

We performed a series of \FLASH simulations changing the
masses, concentration parameters, impact parameter, and infall
velocity of the system to find the best model for \CIZSAUSAG.
Although the main goal of our study was not to carry out a systematic 
parameter search, in order to describe qualitatively the effect of 
changing initial parameters, we also show representative 
simulations which do not resemble \CIZSAUSAG.
We list the initial parameters for those simulations
we present in this paper in Table~\ref{T:TABLE1}.
The IDs of our runs are listed in the first column indicating the 
impact parameter ($P$) in kpc and infall velocity ($V$) in 100 \KMSEC of each run
as $PijkVlm$.
The virial masses (in units of \MSUNFOUR) and concentration parameters
of the two subclusters are shown in columns 2-5.
In columns 6 and 7 we list the impact parameters in kpc and infall 
velocities in \KMSEC.

%
%
\begin{deluxetable}{ccccccc}[t]
\tablecolumns{11}
\tablecaption{                       \label{T:TABLE1} 
 IDs and input parameters for different models used in our hydrodynamical simulations.
} 
\tablewidth{0pt} 
\tablehead{ 
 \multicolumn{1}{c}                {ID\,\tablenotemark{a}}       &
 \multicolumn{1}{c}   {M$_{vir1}$\,\tablenotemark{b}}     &
 \multicolumn{1}{c}   {c$_{vir1}$\tablenotemark{b}}        &
 \multicolumn{1}{c}   {M$_{vir2}$\,\tablenotemark{c}}          &
 \multicolumn{1}{c}   {c$_{vir2}$\tablenotemark{c}}        &
 \multicolumn{1}{c}              {P\,\tablenotemark{d}}          &
 \multicolumn{1}{c}   {V$_{in}$\,\tablenotemark{e}}       
 }
 \startdata  
   P120V25    &   5.0    &   8     &   3.9   &  8  &  120  &  2500   \\ \hline
   P050V25    &   5.0    &   8     &   3.9   &  8  &    50  &  2500   \\ \hline
   P180V25    &   5.0    &   8     &   3.9   &  8  &  180  &  2500   \\ \hline
   P100V25    &   5.0    &   8     &   3.0   &  8  &  100  &  2500   \\ \hline
   P150V20    &   5.0    &   5     &   4.5   &  6  &  150  &  2000   \\ \hline
   P150V30    &   4.0    &   8     &   3.5   &  8  &  150  &  3000   \\ \hline
   P200V18    &   16.0  &   6     &   6.0   &  8  &  200  &  1800
 \enddata
\tablenotetext{a}{ID of the runs indicate  the impact parameters in kpc and infalling velocities in 100 \KMSEC
                            and the mass o fthe main cluster in \MSUNFOUR.}
\tablenotetext{b}{Virial mass  in \MSUNFOUR and concentration parameter for the main cluster (1).}
\tablenotetext{c}{Virial mass in \MSUNFOUR and concentration parameter for the infalling cluster (2).}
\tablenotetext{d}{Impact parameter, P, in units of kpc.}
\tablenotetext{e}{Infall velocity of cluster 2, V$_{in}$, in \KMSEC.
\vspace{.4 cm}}
\end{deluxetable}  

\section{Results and discussion}
\label{S:RESULTS}


\subsection{Deprojecting \CIZSAUSAG}
\label{SS:DEPROJECT}

We used the X-ray morphology, the positions of the merging shocks
\citep{OgreanET2014,AkamatsuET2015,WeerenET2010Sci}, 
and the positions of the dark matter centers of the two subclusters 
derived from weak lensing measurements of 
\CIZSAUSAG \citep{OkabeET2015}, 
to constrain the masses, the concentration parameters,
the impact parameter, and the infall velocity of the system.

After each run, 
we rotated the system with an angle, $\theta$, out of the plane 
of the sky (assumed to coincide with the main
plane of the collision containing the two cluster centers and the 
relative velocity vector) to match the projected distance 
with that of the observed \citep{OkabeET2015}.
The second rotation around the
axis (``roll angle'') connecting the two dark matter centers with an angle, $\varphi$,
was constrained by the observed
X-ray morphology \citep{OgreanET2014}. We choose those outputs
(epochs) which could be rotated in a way that the two merging shocks could be
projected to match the observed positions \citep{AkamatsuET2015,OgreanET2014}.
The shocks were located based on projected pressure gradients, 
a detailed description of our method to generate the X-ray and mass surface density 
maps can be found in \cite{MolnarBroadhurst2015}.

We display images of CIZA J2242.8+5301 based on multi-wavelength observations 
and our best model in Figure~\ref{F:XRAYDMSHOCK}.
The first panel (from the left) shows an image from \CHANDRA observations 
(Figure 1 from \citealt{OgreanET2014}) overlaid the shock positions based on \SUZAKU
observations proposed by \cite{AkamatsuET2015} (yellow lines).
Comparing the black and gray data points of the temperature around the 
southern relic on the left panel of Figure 7 of \cite{AkamatsuET2015},
we suggest that the southern shock is farther out 
than their estimate, it is located at the outer edge of the relics, 
shown as light green line in our figure (Figure~\ref{F:XRAYDMSHOCK}).
The second panel displays a Subaru image (gi color)
with the galaxy number density contours (white contours), 
the XMM-Newton X-ray luminosity distribution (red contours), 
and the WSRT radio emission (green contours) 
overlaid (Figure 1 from \citealt{DawsonET2015}).
We show our best model for \CIZSAUSAG, 
which has a total virial mass of M$_{vir} = 8.9 \times$\MSUNFOUR, 
a mass ratio of 1.3:1, an impact parameter of P = 120 kpc,
and an infall velocity, V$_{in} = 2500$ \KMSEC
(with rotations: $\theta = 15\degree$ and $\varphi = 20\degree$)
in the third panel of Figure~\ref{F:XRAYDMSHOCK} (run P120V25).
This model has a similar elongated -- tidally stretched -- X-ray morphology as \CIZSAUSAG , 
the positions of the dark matter peaks and the shocks also match the 
observations.

In Figure~\ref{F:XRAYBAD}, we show models which do not fully resemble \CIZSAUSAG,
but deviate significantly in one or more ways from the body of constraining data. 
The panels in the first row show models with the same initial parameters as
our best model, but with an impact parameter that is either too small (50 kpc) or too large 
(180 kpc), resulting in a bulky X-ray core or a second peak in the north, respectively.
In the second and the third row we show models,
which have two large or too small mass ratios, or a very high infall
velocity with a small mass ratio which produces the southern shock tilted more to the west
than the observed one (1st panel in the third row).
Our model with a more massive main cluster 
(the total mass about the same as the one derived by \citealt{OkabeET2015},
but with somewhat higher mass ratio, 2.66:1 vs. 2:1)
show a thick X-ray bridge between the two dark matter centers not a thin, 
elongated feature as in \CIZSAUSAG (2nd panel in the third row).
For the input parameters of all models see Table~\ref{T:TABLE1}.
In general, very massive merging clusters 
(M$_{vir} \sim 10^{\,15}\,$\ensuremath{\mbox{\rm M}_{\odot}})
keep their gas and no thin tidal bridge would form between the two dark matter centers,
and thus they do not look like \CIZSAUSAG.

\subsection{Properties of merging shocks in \CIZSAUSAG}
\label{SS:SHOCKPROPERTIES}

Our simulations clearly demonstrate that 
the merging cluster, \CIZSAUSAG, is being seen just after the first core passage
(in agreement with \citealt{WeerenET2011MNRAS418}) and before 
any subsequent core passages.
The less massive infalling cluster, moving north, has passed the core of the main cluster 
and driving forward a bow shock currently located at the top of the images of the cluster 
(Figure~\ref{F:XRAYDMSHOCK}).
We predict a back shock propagating south, in the opposite direction 
to the infallen northern cluster. The two shocks are not expected to be symmetrical 
due to the different sizes of the merging clusters.
Our simulations suggest that the northern subcluster is less massive 
in agreement with \cite{OkabeET2015}, and differing in conclusion with 
\cite{WeerenET2011MNRAS418}. who suggest that the northern 
subcluster is more massive based on the argument that it has larger 
relics.

The pre-shock gas that lies ahead of the bow shock (the northern shock;
to the right in Figure~\ref{F:TEMPERPROFSUZ})
is driven northwards by the gas pressure of the infalling cluster, 
but belongs to the main cluster. 
The pre-shocked gas behind the opposite shock (the southern shock) 
belongs to the infalling cluster and this back shock is moving faster relative 
to the pre-shock gas than the bow shock, and has a higher temperature. 
The position of this southern shock is depicted by 
the solid line in Figure~\ref{F:TEMPERPROFSUZ}).

The Mach number derived from the temperature jump at the shock in X-ray 
observations based on the Rankine-Hugoniot jump conditions
is given by:
\begin{equation}
   \frac{T_2}{T_1} = \frac{5 \MACH^2 + 14 \MACH^2 -3}{16 \MACH^2}
,
\label{E:MACHT2T1}
\end{equation}
where $T_1$ and $T_2$ are the pre- and post-shock temperatures,
and \MMACH is the Mach number (e.g., \citealt{BotteonET2016,AkamatsuET2015};
for a review see \citealt{MarkevitchVikhlinin2007}).
Using the temperature jump from our simulations at the shocks 
in Equation~\ref{E:MACHT2T1}, 
we obtain Mach numbers for the northern and southern shocks
(the bow shock and back shock):
$\MACH_{n,simu} = 6.5$ and $\MACH_{s,simu} = 7.8$.
Based on our set of simulations, we find that, in general, 
the back shocks are stronger than the bow shocks
in binary merging clusters having a larger temperature jump.

Our simulations predict spatially smooth
shock fronts for the bow shock and the back shock
(the northern and southern shocks as shown in Figure~\ref{F:XRAYDMSHOCK}).
For CIZA J2242.8+5301, the morphology of the radio emission associated with the 
bow shock is smooth, and very similar in curvature as our simulations predict. 
In the south the angular location of our predicted shock front is coincident with 
our predictions but always lies at a somewhat larger radius. 
Interestingly, the observed X-ray shock front is also claimed to lie at somewhat larger 
radius than the  southern radio relics, although the precision of this position could be 
improved with deeper X-ray data. We also note that the observed radio structures that 
lie close the predicted back shock do not form a continuous sharp arc, but appears 
bifurcated with a more irregular morphology. 
A possible reason for this imperfect correspondence in the radio might be that the 
backward shock is impeded by subsequently infalling gas associated with the same
 filament that was connected to the infalling northern cluster, 
The next wave of shocks collides that has has impeded and brakes up 
producing patchy relics that lag behind the predicted location 
of our ideal simulations.

\subsection{Bias in Mach numbers derived from X-ray observations}
\label{SS:BIASMACH}

Some recent observations of merging
clusters of galaxies show that the Mach numbers of merging shocks
based on radio observations are about twice as large as 
those derived from X-ray observations 
(e.g., A2255: 
$\MACH_{radio} = 2.7$, $\MACH_{xray} = 1.4$, \citealt{AkamatsuET20161203058};     
RX J0603.3+4214 (the ``Toothbrush'' cluster): 
$\MACH_{radio} = 2.8$, $\MACH_{xray} = 1.2$, \citealt{WeerenET2016ApJ818};
the northern shock in \CIZSAUSAG (the ``Sausage'' cluster):
$\MACH_{radio} = 4.6$, $\MACH_{xray} = 2.7$,
\citealt{WeerenET2010Sci,AkamatsuET2015}).

However, the physical gas temperature jump across the shock 
should be used in Equation~\ref{E:MACHT2T1}, which is not observed.
Only the projected temperature, which is the LOS integrated temperature 
weighted by the emission measure and convolved 
with the response function of the X-ray detector, 
can be derived from X-ray observations directly
(i.e., without assuming a model for the LOS distribution of the gas).
The usual method is to assume a simple geometry of
the shock and use that model to deproject the observed image and
derive the temperature jump (e.g., \citealt{MenanteauET2012}).
Some corrections are used occasionally to deal with the fall of the temperature 
with distance from the cluster center 
of the undisturbed (pre-shocked) cluster gas (e.g., \citealt{AkamatsuET2015}).

In principle, the best way to deproject a merging cluster, 
and derive the Mach numbers for the shocks is to model the system 
using a full N-body/hydrodynamical code.
We demonstrate the power of a full numerical 
simulation by comparing the Mach number we 
derived using the best model 
to that inferred from \SUZAKU observations of \CIZSAUSAG 
\citep{AkamatsuET2015}.
The derivation of Mach numbers from \SUZAKU observations 
is difficult, because, in addition to the above mentioned
problems with deprojection, it also has a low angular
resolution of 1.\arcmin6.
Also, the shocked gas can reach a temperature of 20-50 keV,
but the effective area of \SUZAKU's X-ray Imaging Spectrometer (XIS)
cuts off at about $\sim8$ keV and exponentially small for gas temperatures higher
than 10 keV.
Note, that this is also a problem for \CHANDRA and \XMM.

In Figure~\ref{F:TEMPERSUZ} we display temperature maps based on 
our best model for \CIZSAUSAG (run P120V25).
The left panel shows the 
physical temperature (2D cut through the main plane of the collision),
the middle panel shows a temperature map taking into account projection effects 
(spectroscopic temperature of \citealt{MazzottaET2004}), 
and the right panel displays a mock \SUZAKU observation 
(spectroscopic temperature convolved with the point spread function, PSF of \SUZAKU).
From this figure we can see that projection effects and a convolution
with a low resolution PSF soften the shocks and they seem to be closer 
to the center of the merging clusters.

%
%
\begin{figure}[t]
\includegraphics[width=.477\textwidth]{\FIGURES/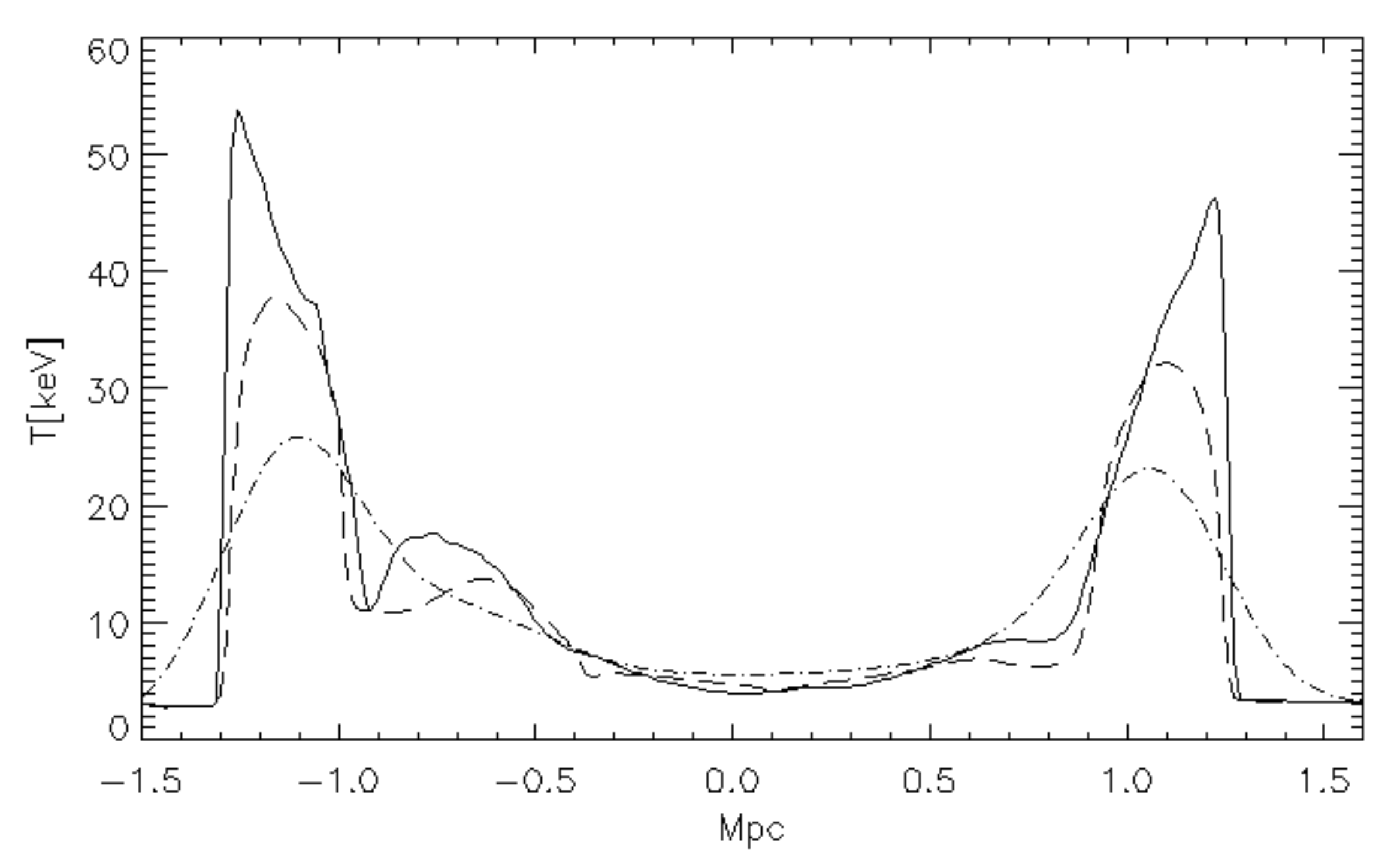}
\caption{
Temperature profiles across the bow shock (northern shock) 
and the back shock (southern shock)
associated with the infalling and main cluster after the first core passage
based on our best model (from Figure~\ref{F:TEMPERSUZ}).
Solid, dashed and dash-dotted lines represent the physical gas temperature,
the spectroscopic temperature and a mock \SUZAKU observation.
\label{F:TEMPERPROFSUZ}
}
\end{figure} 

Figure~\ref{F:TEMPERPROFSUZ} shows the 
temperature profile across the bow shock and the back shock
associated with the infalling and main cluster after the first core passage.
Solid, dashed and dash-dotted lines represent the physical gas temperature,
the predicted spectroscopic temperature \citep{MazzottaET2004},
and a mock \SUZAKU observation
based on the spectroscopic temperature
(the spectroscopic temperature convolved with the PSF of \SUZAKU.)
From this Figure, we can see that, due to projection
effects and the low angular resolution, the measured
temperature jump at the shocks by \SUZAKU would be biased low
by a factor of two 
(compare the solid and dash-dotted temperature profiles).
If $T_2$ is underestimated by a factor of two, the
Mach number, \MMACH, derived from the temperature jump
is going to be biased low by $\Delta \MACH \sim 2$.

\cite{AkamatsuET2015}, for example, measured 
a Mach number of $\MACH_{n,A} = 2.7_{-4}^{+7}$
at the northern shock (bow shock) using \SUZAKU observations.
Our N-body/\-hydro\-dynamical simulations suggest a correction 
of about  $\Delta \MACH \sim +2$ for \SUZAKU observations.
With this correction,
the Mach number for the northern shock in \CIZSAUSAG
would be $\MACH_{n,A,corrected} \sim 4.7$.
This corrected Mach number for the northern shock agrees well with 
$\MACH_{n,radio} = 4.6_{-0.9}^{+1.3}$, the value derived from 
radio observations by \cite{WeerenET2010Sci}.

%
%
\begin{figure}[t]
\includegraphics[width=.477\textwidth]{\FIGURES/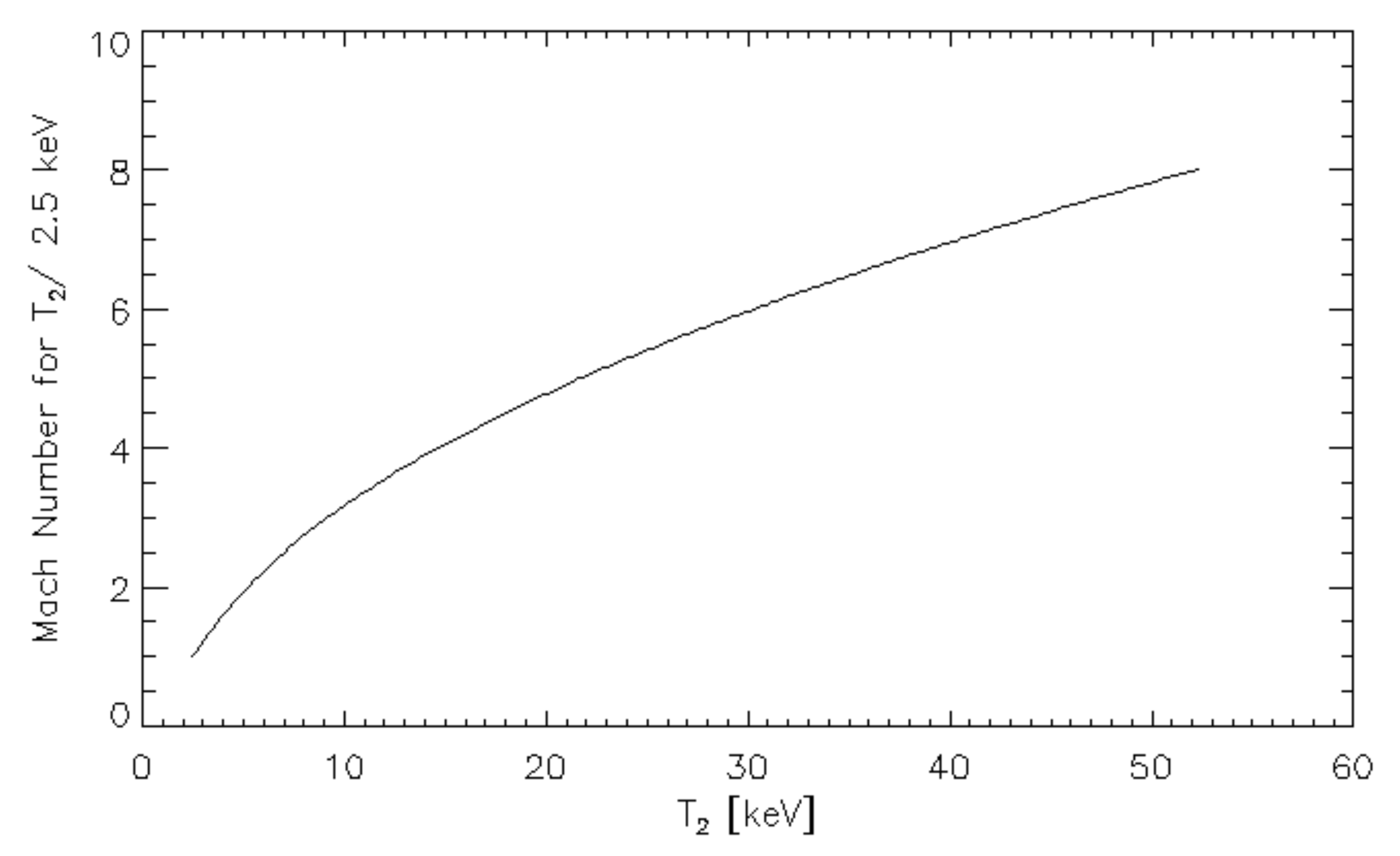}
\caption{
The Mach number as a function of post-shock temperature, $T_2$,
holding the pre-shock temperature fixed at a value
expected in the outer parts of a cluster, $T_1 = 2.5$ keV.
If $T_2$ is underestimated by a factor of two, the derived Mach number, \MMACH, 
is biased low by $\Delta \MACH \sim 2$.
\vspace{0.3 cm}
\label{F:MACHT1T2}
}
\end{figure} 

However, averaging the projected temperature on a larger area in the sky
further lowers the measured shock temperature, and thus 
the Mach numbers for the bow shock (northern) and the back shock (southern)
derived by \cite{AkamatsuET2015}, 
$\MACH_n = 2.7_{-0.4}^{+0.7}$ and $\MACH_s = 1.7_{-0.3}^{+0.4}$, 
are most likely biased even lower.
Therefore, even with our suggested correction, $\MACH_{n,A,corrected} \sim 4.7$,
we may underestimate the Mach number for the northern shock, $\MACH_{n}$.
Note that using the temperature jump from our simulations, we obtain
$\MACH_{n,simu} = 6.5$ and $\MACH_{s,simu} = 7.8$ 
(see Section~\ref{SS:SHOCKPROPERTIES}).

We illustrate the bias in the Mach number due to measurement
errors in the post-shock temperature in Figure~\ref{F:MACHT1T2}.
In this figure, using Equation~\ref{E:MACHT2T1}, 
we show the Mach number as a function of 
post-shock temperature, $T_2$, holding the pre-shock temperature 
fixed at a value expected in the outer parts of a cluster, $T_1 = 2.5$ keV.
This figure suggests that, in general, if 
$T_2$ is underestimated by a factor of two, the Mach number 
is going to be biased low by $\Delta \MACH \sim 2$.

\subsection{Bias in flux measurements of radio relics due to SZ contamination from shocks}
\label{SS:BIASRADIO}

Radio relics, found in the outskirts of clusters of galaxies,
are elongated synchrotron radio sources with a length 
in the order of a Mpc.
The radio emission of relics is highly polarized and have a steep spectrum.
The relativistic electrons emitting the radiation are assumed to 
be accelerated to high energies by shocks in merging clusters
(for a recent review see \citealt{FerettiET2012}).

The main physical mechanism responsible for the 
particle acceleration at cluster merger shocks has not been
identified yet. It is a subject of active research.
The radio flux and the spectral index as a function of distance from the 
shock front provides important constraints 
on particle acceleration models, because they are related to ``spectral aging'' of the 
relativistic electrons due to synchrotron and inverse Compton energy losses.
\CIZSAUSAG, the ``Sausage'' cluster, since in this cluster the projected
distance from the shock gives a good approximation to the physical
distance (the collision is close to the plane of the sky), 
have been used to test different particle acceleration models,
e.g., DSA mechanism (\citealt{StroeET2016},
also analyzed 1RXS J0603.3+4214, the ``Toothbrush'' cluster); 
fossil electrons (accelerated by the DSA mechanism) are 
reaccelerated as the shock runs them over \citep{KangRyu2015ApJ809};
turbulent cosmic ray reacceleration \citep{FujitaET2016}; and 
magnetic turbulence generated by the amplification of the magnetic field
\citep{DonnertET2016}.

DSA acceleration predicts a synchrotron power law spectrum.
However, recent radio observations found curved spectra of some relics, 
which is inconsistent with the prediction of a simple DSA model
(e.g., 1RX J0603.3+4214: \citealt{WeerenET2016ApJ818};
\CIZSAUSAG: \citealt{StroeET2016}).
Among other explanations (listed in the previous paragraph)
it was suggested that around the shock front the thermal SZ effect 
may contaminate the radio flux measurements by lowering the 
measured flux from radio relics, and as a result, the radio spectrum 
becomes curved (e.g., \citealt{BasuET2016AA591}).

%
%
\begin{figure}[t]\includegraphics[width=.477\textwidth]{\FIGURES/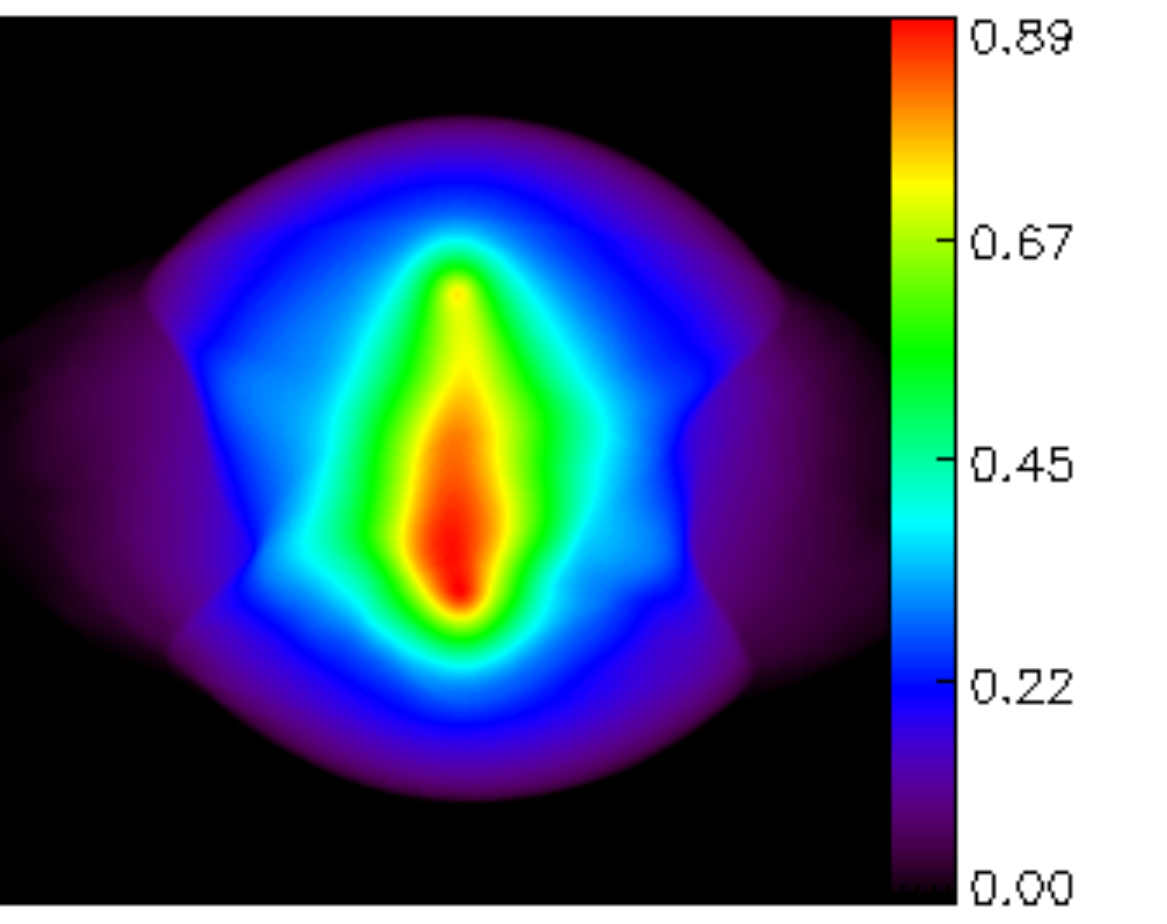}
\caption{
\COMPTONY map of our best model for \CIZSAUSAG (run P120V25).
The figure is 3.1 Mpc a side. The units are \COMPTONY $\times 10^4$.
The infalling cluster passed the main cluster, it is moving north.
The maximum of the \COMPTONY parameter is around the core of the main cluster.
Note that the collision axis is also aligned vertically, as in Figure 3.
\label{F:COMPTONYMAP}
}
\end{figure} 

In Figure~\ref{F:COMPTONYMAP} we show the \COMPTONY map of 
our best model for the merging cluster \CIZSAUSAG (run P120V25).
The units are \COMPTONY $\times 10^4$.
The infalling cluster passed the main cluster, it is located above it, 
and moving upward (north).
The bow shock, ahead of the infalling cluster, moving north
and the back shock moving south, can be clearly seen
(sharp transition from dark blue to magenta).
This predicted \COMPTONY map can be used to test our model with 
future high resolution SZ observations.

\begin{figure}[t]
\includegraphics[width=.477\textwidth]{\FIGURES/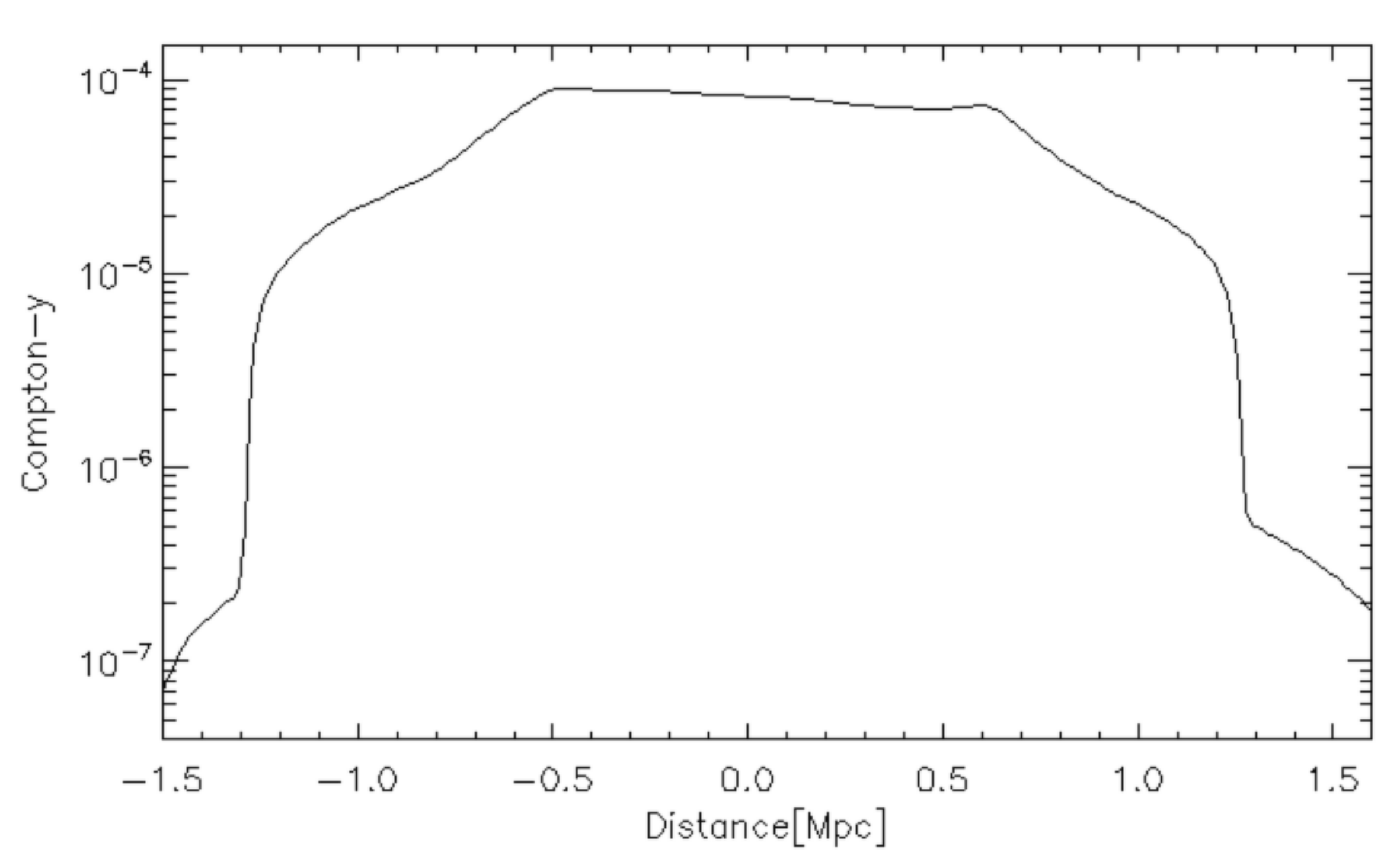}
\caption{
\COMPTONY profile from Figure~\ref{F:COMPTONYMAP}
across the two shocks as a function of distance from the cluster center (in Mpc).
The infalling cluster, moving to the right. passed the core of the main cluster, 
its center is located at about +0.6 Mpc. The center of the main cluster is at -0.5 Mpc.
The sharp drops in the \COMPTONY parameter associated with the bow shock 
(northern shock) at +1.25 Mpc (ahead of the infalling cluster moving to the right) and 
the back shock (southern shock) at -1.3 
(behind the main cluster moving in the left).
\label{F:COMPTONYPROF}
}
\end{figure}

In Figure~\ref{F:COMPTONYPROF} we show the \COMPTONY profile, 
across the two shocks, extracted from the same data shown 
in Figure~\ref{F:COMPTONYMAP}, 
as a function of distance from the cluster center (in Mpc).
The infalling cluster passed the core of the main cluster, its center 
is located at about +0.6 Mpc, it is moving to the right. 
The center of the main cluster is at -0.5 Mpc.
The sharp drop in the \COMPTONY parameter associated with the bow shock 
(northern shock), ahead of the infalling cluster moving to the right, at +1.25 Mpc,
and the back shock (southern shock) 
behind the main cluster moving to the left, at -1.3, can be clearly seen.

%
%

The \COMPTONY parameter drops more than an order 
of magnitude at the merger shocks (see Figure~\ref{F:COMPTONYPROF}).
From the drop in the \COMPTONY parameter, following 
\cite{BasuET2016AA591}, we estimate the decrease in the radio flux density
due to the SZ effect at the shocks using 
\begin{equation}
\left(\dfrac{\Delta S_{\nu}}{\mathrm{mJy/\mathrm{arcmin}^2}}\right) = 
          \dfrac{1}{340} \left(\dfrac{\Delta T_{\mathrm{RJ}}}{\mathrm{mK}}\right) 
	  \left(\dfrac{\nu}{\mathrm{GHz}}\right)^2
,
\label{E:SZFLUX}
\end{equation}
where $\nu$ is the frequency in GHz, 
$\Delta T_{\mathrm{RJ}} = -2 y T_{CMB}$ is the thermal SZ decrement in 
Rayleigh-Jeans radiation temperature in mK, 
and the drop of the radio flux is in mJy/arcmin$^{2}$ (e.g., \citealt{birkinshaw1999}).
We find that that at the northern and southern shocks
at 30 GHz, the radio flux drops 
$\Delta S_n = -0.072$ mJy/arcmin$^2$ and $\Delta S_s = -0.075$ mJy/arcmin$^2$.

\section{Conclusions}
\label{S:CONCLUSIONS}

We have performed a wide set of self-consistent N-body/\-hydro\-dynamical simulations 
(using \FLASH) to seek a representative solution for \CIZSAUSAG, and to help understand 
better the level of deprojection on the Mach numbers of shocks affecting X-ray observations 
of this and other similar clusters.
We have modeled \CIZSAUSAG as a binary merger, constraining the initial
parameters using lensing, X-ray and radio observations. 
The X-ray morphology and the locations of the two lensing centroids we find help 
constrain the impact parameter and the infall velocity. 
The positions of the outgoing shocks were constrained by X-ray and radio observations.

Our numerical simulations represent the 
first attempt to model \CIZSAUSAG using self-consistent 
N-body/\-hydro\-dynamical simulations. 
We can appreciate from these simulations how tidal effects influence the gas 
distribution lying between the two clusters allowing us to identify suitable combinations 
of these initial parameters using the detailed X-ray morphology.  
Note that other models have not been able to benefit from this.
For example, \cite{WeerenET2011MNRAS418} assumed,
for simplicity, fixed non-interacting  shapes for the gravitational potentials
of the two merging clusters moving on pre-calculated paths.

We have demonstrated that low angular resolution X-ray telescopes 
(e.g., \SUZAKU) significantly underestimate the shock temperature
as much as a factor of two, and thus the resulting Mach number may 
be biased low by $\Delta \MACH \sim 2$.
Adding this correction to the \SUZAKU result for the
Mach number of the northern shock,
$\MACH_{n,A} \sim 2.7$ \citep{AkamatsuET2015}, 
we obtain $\MACH_{n,A,corrected} \sim 4.7$, 
which agrees well with the result from radio observations,
$\MACH_{n,radio} = 4.6$ \citep{WeerenET2010Sci}.

We have suggested that the relics around the 
northern and southern shocks in \CIZSAUSAG look different,
the southern relics, in contrast to our simulations, are patchy and
irregular in shape, perhaps because of a filament of gas that 
follows the infalling northern cluster has impeded 
and disturbed the back shock on its way out towards the south
colliding with the subsequently infalling filament of gas.

The main mechanism for particle acceleration around shocks
in merging galaxy clusters is an active subject of research.
The shape of the radio spectrum from radio relics located around shocks
provides constraints on particle acceleration models associated with shocks.
However, the thermal SZ effect associated with the dense shocked gas may 
act to significantly lower the measured radio flux at shocks, as a function of frequency. 
We have simulated \COMPTONY maps based on our best model for \CIZSAUSAG,
and shown that the drops in the \COMPTONY parameter 
due to the merging shocks may reach more than one order of magnitude.
Our model indicates that at the northern and southern shocks at 30 GHz, 
the radio flux drops -0.072 mJy/arcmin$^2$ and -0.075 mJy/arcmin$^2$.
This sharp discontinuity, however, is smoothed out by the PSF of the 
observing radio telescopes. In agreement with \cite{BasuET2016AA591},
we find that merging shocks may considerably reduce the measured flux 
of radio relics depending mainly on the masses and 
relative velocities of the merging clusters.

\acknowledgements
We thank Huub R{\"o}ttgering for discussions on the interpretation of the radio data.
The code \FLASH\ used in this work was in part developed by the
DOE-supported ASC/Alliance Center for Astrophysical Thermonuclear
Flashes at the University of Chicago.  
We thank the Theoretical Institute for Advanced Research in Astrophysics,
Academia Sinica, for allowing us to use its high performance computer facility 
for our simulations.
This research has made use of the NASA/IPAC
Extragalactic Database (NED) which is operated by the Jet Propulsion
Laboratory, California Institute of Technology, under contract with
the National Aeronautics and Space Administration.


%
%
\bibliographystyle{apj}


\end{document}